\documentclass{cimento}
\usepackage{graphicx}
\title{The GRB of 1999 January 23: prompt emission and broad-band afterglow modeling}
\author{A. Corsi\from{ins:1}\from{ins:2},L. Piro\from{ins:1},E. Kuulkers\from{ins:3},L. Amati\from{ins:4},L.A. Antonelli\from{ins:5},E. Costa\from{ins:1},\\M. Feroci\from{ins:1}, F. Frontera\from{ins:4}\from{ins:6},C. Guidorzi\from{ins:6},J. Heise\from{ins:7}, J. in't Zand\from{ins:7},\\E. Maiorano\from{ins:4}\from{ins:8},E. Montanari\from{ins:6},L. Nicastro\from{ins:9},E. Pian\from{ins:10}\atque P. Soffitta\from{ins:1}.}
\instlist{\inst{ins:1}IASF-CNR, Via Fosso del Cavaliere 100, I-00133 Rome, Italy.\inst{ins:2}University ``La Sapienza'', Piazzale Aldo Moro 5, I-00185 Rome, Italy.
\inst{ins:3}ESA-ESTEC, Science Operations \& Data Systems Division, SCI-SDG, Keplerlaan 1, 2201 AZ Noordwijk, The Netherlands.
\inst{ins:4}IASF-CNR, Via Gobetti 101, I-40129 Bologna, Italy.
\inst{ins:5}Rome Astronomical Observatory, Via di Frascati 33, I-00044 Rome, Italy.
\inst{ins:6}Physics Department, University of Ferrara, Via Paradiso 11, I-00044 Rome, Italy.
\inst{ins:7}Space Research Organization Netherlands, Sorbonnelaan 2, 3584 CA Utrecht, The Netherlands.\inst{ins:8}Astronomy Department, University of Bologna, Via Ranzani1, I-40126 Bologna, Italy.
\inst{ins:9}IASF-CNR, Via Ugo la Malfa 153, I-90146 Palermo, Italy.
\inst{ins:10}INAF, Osservatorio Astronomico di Trieste, Via G.B. Tiepolo, 11 - I-34131 Trieste, Italy.}
\begin{document}
\shorttitle{The GRB of 1999 January 23}
\shortauthor{A. Corsi, L. Piro, E. Kuulkers et al.}
\PACSes{\PACSit{95.75.Fg}{Spectroscopy and spectrophotometry}\PACSit{95.85.Nv}{X-ray}\PACSit{98.70.Rz}{gamma-ray sources; gamma-ray bursts}}
\maketitle
\begin{abstract}
We report on \textit{Beppo}SAX simultaneous X- and $\gamma$-ray observations of the bright $\gamma$-ray burst (GRB) 990123. We present the broad-band spectrum of the prompt emission, including optical, X- and $\gamma$-rays, confirming the suggestion that the emission mechanisms at low and high frequencies must have different physical origins. We discuss the X-ray afterglow observed by the Narrow Field Instruments (NFIs) on board \textit{Beppo}SAX and its hard X-ray emission up to $60$~keV several hours after the burst, in the framework of the standard fireball model. The effects of including an important contribution of Inverse Compton scattering or modifying the hydrodynamics are studied. 
\end{abstract}
\section{Prompt emission}
\begin{figure}[t]
\begin{center}
\includegraphics[width=0.40\textwidth,angle=-90]{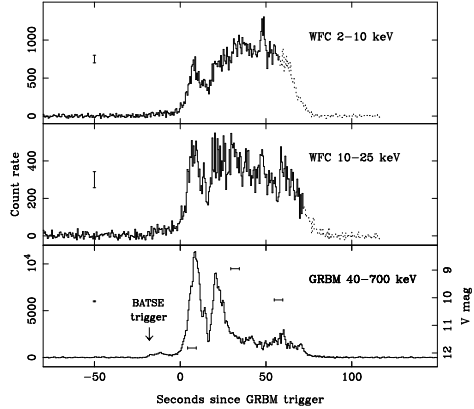}
\caption{Prompt burst profile of GRB~990123 at X-ray 
(WFC: 2--10~keV and 10--25~keV)
and at $\gamma$-ray (GRBM: 40--700~keV) energies. Count rate (counts s$^{-1}$)
is given as a function of time after the GRBM trigger. The dotted line refers to the part where more than 10\%\ of the 
intensity is lost due to Earth-atmospheric absorption. Typical error bars are given in the left part of the panels; the times of the BATSE
trigger and of the prompt optical measurements 
\cite{Akerlof1999} in the bottom panel.}
\label{figura1}
\end{center}
\end{figure}

\begin{figure}
\begin{center}
\includegraphics[width=0.40\textwidth,angle=-90]{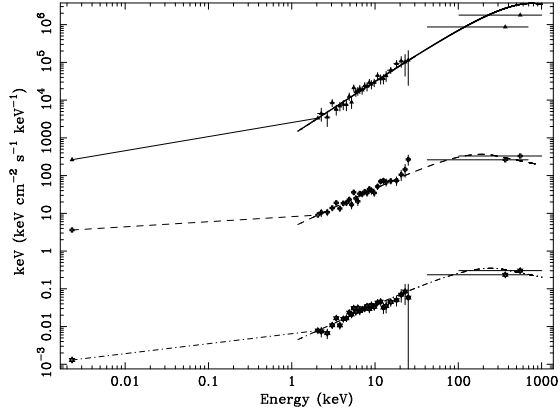}
\caption{
Simultaneous multi-wavelength spectra derived at three times during the burst.
The ROTSE data points \cite{Akerlof1999} have been connected with a line to the best fit model at $2$ keV to guide the eye (solid line for the first ROTSE observation, dashed and dashed-dotted lines for the second and third ones, respectively). The data relative to the first (third) time interval have been shifted up (down) of a factor $1000$, respectively.}
\label{fig2}
\end{center}
\end{figure}

GRB~990123 was one of the brightest $\gamma$-ray bursts detected by the \textit{Beppo}SAX satellite; the Gamma-Ray Burst Monitor (GRBM) was triggered by GRB~990123 on 1999 January 23.40780 UT, $\sim 18$~s after the CGRO trigger; the burst was detected for about 100~s and it presented two major pulses, the brighter of which had a peak intensity of $(1.7\pm0.5)\times10^{-5}$~ergs~cm$^{-2}$~s$^{-1}$ between 40--700~keV (Fig. \ref{figura1}). GRB 990123 was also associated with the first observation of a prompt optical emission \cite{Akerlof1999}, of a short lived radio emission \cite{Kulkarni1999b} and of a clear break in the optical afterglow light curve \cite{Castro-Tirado1999,Fruchter1999,Holland2000,Kulkarni1999a}.
The total fluence between $40$~keV and $700$~keV was $(1.9\pm0.2)\times10^{-4}$~ergs~cm$^{-2}$; this value improves the one given in the paper by Amati et \textit{al.} \cite{Amati2002}. For the WFC and GRBM data during the first three ROTSE exposures we find a spectral slope of $-0.57\pm0.06, -0.95\pm0.09, -1.0\pm0.1$, respectively. The extrapolation of the high energy time resolved spectra to optical frequencies falls at least 2 orders of magnitude below the simultaneous optical measurements (Fig. \ref{fig2}), suggesting different physical origins for the emission mechanism at low and high frequencies and confirming the idea of a reverse shock origin for GRB~990123 optical flash \cite{Briggs1999,Galama1999,Sari1999}. A more detailed description of the WFC and GRBM data analysis of the prompt emission and of the broad-band afterglow modeling presented in the following section, is given in Corsi et \textit{al.} \cite{Corsi2004}.
\section{Broad-band afterglow modeling}
The \textit{Beppo}SAX follow-up observation lasted from 1999 January $23.65$~UT until January $24.75$~UT. The source was detected by the LECS and MECS units, and also observed by the PDS in the 15--60~keV energy band. The 2--10~keV flux decreases with time as a power law of index $\alpha_{\rm X}=-1.46\pm0.04$; see the paper by Maiorano et \textit{al.} \cite{Maiorano2004} for a complete description of the WFC, MECS and PDS light curves in the 2--10 keV and 15--28 keV energy bands. During the first day, the optical flux in the Gunn r band decreases with $\alpha_{\rm opt}=-1.10\pm0.03$ \cite{Holland2000,Kulkarni1999a}. For the LECS (0.1--2~keV), MECS (2--10~keV) and PDS (15--60~keV) data during the first $20$~ks, we find a best fit spectral index of $\beta_{\rm X}=-0.82\pm0.10$. Different values for the optical-to-IR spectral index before the time of the break have been reported; those values are between $-0.8\pm 0.1$ \cite{Kulkarni1999a} and $-0.60\pm0.04$ \cite{Maiorano2004}. In our following discussion, we consider the mean optical-IR observed spectral index of $-0.75 \pm 0.068$, as derived by Holland et \textit{al.} \cite{Holland2000}, and address for the case $\beta_{\rm opt}=-0.60\pm0.04$ \cite{Maiorano2004} in the final part.

\begin{figure}
	\begin{center}
		\includegraphics[width=0.40\textwidth,angle=-90]{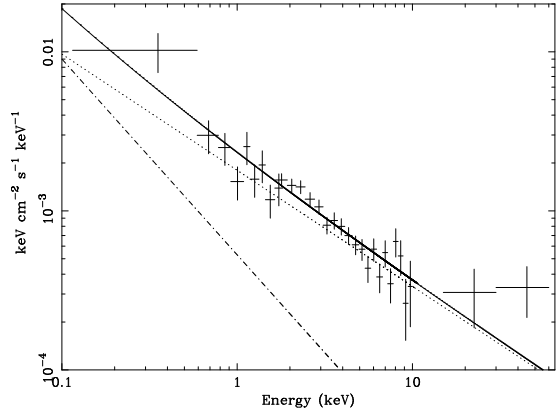}
		\caption{GRB 990123 X-ray spectrum during the first $20$~ks of the NFIs observation. The model spectrum is the sum of two power-law components: one represents the synchrotron contribution and has a fixed spectral index $\beta_{\rm X}=-1.23$, the other accounts for IC scattering and has a fixed spectral index $\beta^{\rm IC}_{\rm X}=-0.73$.}
\label{figura3}
	\end{center}
\end{figure}

In the standard synchrotron fireball model, the closure relationships are: if $\nu_{\rm opt}<\nu_{\rm X}<\nu_{c}$, $\alpha_{\rm opt}-\frac{3}{2}\beta_{\rm opt}=0$, $\alpha_{\rm X}-\frac{3}{2}\beta_{\rm X}=0$, $\alpha_{\rm opt}-\alpha_{\rm X}=0$, $\beta_{\rm opt}-\beta_{\rm X}=0$; if $\nu_{\rm opt}<\nu_{\rm c}<\nu_{X}$, $\alpha_{\rm opt}-\frac{3}{2}\beta_{\rm opt}=0$, $\alpha_{\rm X}-\frac{3}{2}\beta_{\rm X}-\frac{1}{2}=0$, $\alpha_{\rm opt}-\alpha_{\rm X}-\frac{1}{4}=0$, $\beta_{\rm opt}-\beta_{\rm X}-\frac{1}{2}=0$. The corresponding values for $\beta_{\rm opt}=-0.75\pm0.068$, $\beta_{\rm X}=-0.82\pm0.10$, $\alpha_{\rm opt}= -1.1\pm0.03$ and $\alpha_{\rm X}= -1.46\pm0.04$ are: $0.02\pm0.11$, $-0.23\pm0.15$, $0.360\pm0.050$, $0.07\pm0.12$; $0.02\pm0.11$, $-0.73\pm0.15$, $0.11\pm0.050$, $-0.43\pm0.12$, respectively.
For $\nu_{\rm opt}<\nu_{\rm X}<\nu_{c}$ the relationship involving the temporal indices is not satisfied; for $\nu_{\rm opt}<\nu_{c}<\nu_{\rm X}$, those involving $\beta_{\rm X}$ are not, giving evidence for a too flat X-ray spectrum. It is then difficult to interpret GRB~990123 afterglow within the basic synchrotron model. We thus consider a model in which the X-ray emission is dominated by IC-scattering of lower energy photons, while the optical one is synchrotron dominated. In the slow-cooling IC dominated phase \cite{Sari2001}, with $\nu_{\rm opt}<\nu_{c}<\nu_{\rm X}$ and $\nu_{\rm X}>\nu^{\rm IC}_{m}$, the closure relations are: $\alpha_{\rm opt}-\frac{3}{2}\beta_{\rm opt}=0$, $\alpha_{\rm X}-\frac{9}{4}\beta_{\rm X}-\frac{1}{4}=0$, $\alpha_{\rm opt}-\frac{2}{3}\alpha_{\rm X}+\frac{1}{6}=0$, $\beta_{\rm opt}-\beta_{\rm X}=0$, all verified within $1\sigma$ ($-0.02\pm0.11$, $0.13\pm0.23$, $0.040\pm0.040$, $0.07\pm0.12$).
Using $p=2.47\pm0.02$ (as derived from $\alpha_{\rm opt}=-1.10\pm0.03$) we obtain $\beta_{\rm X}=-1.23\pm0.02$, $\beta^{\rm IC}_{\rm X}=-0.73\pm0.02$, for the synchrotron and IC component spectral indices, respectively. After having expressed the parameters of the fireball model as functions of $\nu_{c}$, $F_{2\rm keV}^{\rm syn}$, $\nu_{m}^{\rm IC}$, $F^{\rm IC}_{m}$, by minimizing the energy requirement and fitting the 0.1--60 keV spectrum with a synchrotron+IC spectrum, we get: $\nu_{c}\sim2$~eV, $F_{2\rm keV}^{\rm syn}\sim10^{-2}~\mu$Jy, $\nu_{m}^{\rm IC}\sim2~\rm keV$, $F^{\rm IC}_{m}\sim0.83~\mu$Jy (Fig. \ref{figura3}) and $\epsilon_{B}\sim10^{-10}$, $\epsilon_{e}\sim10^{-1}$, $E_{52}\sim7\times10^{4}$, $n_{1}\sim10^{2}$.
We estimate the jet opening angle $\theta_{\rm j}$ \cite{Frail2001} using $t_{\rm j}\sim 2$~d, and obtain $\theta_{\rm j}=0.064\sim3.7^{\circ}$; this implies $E_{\rm j}\sim10^{54}$~ergs for the energy in the jet. A value of the order of $10^{4}$ for $E_{52}$ has recently been proposed by Panaitescu et \textit{al.} \cite{Panaitescu2004}. Finally, we consider the consistency of the model with the $8.46$~GHz data \cite{Kulkarni1999b}; we have:
$F_{8.46\rm GHz}(1~{\rm d})\sim7~\rm mJy$, about a factor of $30$ above the observations, that are lower than $260$~$\mu$Jy \cite{Kulkarni1999b}. Thus, the radio upper limits are not satisfied. As an alternative different approach, we leave unchanged the emission mechanism but change the hydrodynamics: if $\nu_{c}\sim1$ keV, the optical to X-ray spectral index $\beta_{\rm X-opt}=-0.54\pm0.02$ \cite{Kulkarni1999a} implies $p=2.08\pm0.04$, $\beta_{\rm opt}=-0.54\pm0.02$ and $\beta_{\rm X}=-1.04\pm0.02$, that agree at the $\sim 1\sigma$ level with the spectral indices found by Maiorano et \textit{al.} \cite{Maiorano2004}. To account for the temporal decay indices, we change the hydrodynamics by letting $E\propto t^{\delta _{E}}$ and $n\propto t^{\delta_{n}}$ ($t$ is the
observer time) and obtain $\delta_E\sim-0.39$, so the shock should lose energy, and $\delta_n\sim0.41$. These indices give $R(t)\sim t^{0.05}$, where $R$ is the radius of the shock, that implies $n\propto R^8$. Thus, in this alternative scenario, the shock should lose energy while the density should increase rapidly with radius.

\end{document}